\def   \ni {\noindent}

\def   \ssk {\vskip  5truept}

\def   \bsk {\vskip 10truept}

\def   \newline {\hfil\break}

\documentstyle[epsfig]{article}
\begin{document}

\hsize 5truein

\font\abstract=cmr8
\font\keywords=cmr8
\font\caption=cmr8
\font\references=cmr8
\font\text=cmr10
\font\affiliation=cmssi10
\font\author=cmss10
\font\mc=cmss8
\font\title=cmssbx10 scaled\magstep2
\font\alcit=cmti7 scaled\magstephalf
\font\alcin=cmr6 
\font\ita=cmti8
\font\mma=cmr8
\def\ref{\par\noindent\hangindent 15pt}
\null


\title{\ni The Extreme Universe: Some Views From Here}

\bsk \bsk
\author{\ni Elihu Boldt}                                                       
\bsk
\affiliation{\\ Laboratory for High Energy Astrophysics\\
 NASA Goddard Space Flight Center\\
 Greenbelt, MD 20771, USA}
\bsk
\baselineskip = 12pt

\abstract{ABSTRACT \ni \ 
Forty years have passed since the first Explorer orbiting observatory -
the 1958 mission used to discover the Van Allen radiation belts outside the
atmosphere - ushered in the modern age of space science.  Even though in
situ observations of outer space are still restricted to measurements made
within the solar system, we now have access to a wide range of cosmic
signals, extending from the well understood microwave photons indicative of
the earliest epoch of the universe to those apparently inexplicable
ultra-high energy extragalactic cosmic ray particles that are too energetic
(up to $50$ Joules/each) to have survived passage through a cosmological
extent of the pervasive thermal relic radiation field.  In this lecture the
extremes of cosmic ray physics are discussed within the context of
particles having the lowest energy (down to $\sim 10^3$ eV/nucleon) and highest
energy ($>10^{20}$ eV), emphasizing those aspects of astronomy, particularly
gamma-ray and x-ray, that appear to be especially revealing for these
regimes.  }                                                    
\bsk
\baselineskip = 12pt
\keywords{\ni KEY WORDS: cosmic rays - gamma rays - x-rays - black holes }               

\bsk
\baselineskip = 12pt


\text{\ni 1. INTRODUCTION}
\ssk
\ni     

	Over eighty years ago Victor Hess, standing in a balloon-borne gondola,
used simple electroscopes to discover ionizing radiation coming from the
residual atmosphere above him, thereby initiating the rich fields of high
energy physics and high energy astrophysics.  Now we are ready to begin the
next century with major space-borne astronomical observatories for cosmic
x-rays and gamma-rays with powerfully instrumented sophisticated missions
such as AXAF, XMM, Astro-E and INTEGRAL.  In this lecture I trace the
evolution of high energy astrophysics within the context of cosmic X and
gamma radiation, emphasizing those aspects that are especially relevant to
cosmic ray research.  To a large extent it parallels my own post-graduate
involvement in the field over the last four decades and reflects the bias
of my particular interests.

	For the universe traced by baryonic matter, the presently well defined
cosmic X/gamma-ray background has provided a physically critical integral
measure of a vast intervening history, particularly for AGN evolution.  The
implications of this comprehensive measure of accretion powered AGN
activity during all relevant earlier epochs support other recent evidence
for a local abundance of currently inactive supermassive spun-up black hole
quasar remnants, candidate dynamos for the acceleration of cosmic rays.
Finally, I tell of possible space-borne observatories for measuring, from
topside vantage points, the atmospheric fluorescence arising from cosmic
ray initiated air showers and how they could be used to determine the
arrival directions of the very highest energy events, ones that might well
be correlated with those candidate sources associated with supermassive
black hole galactic nuclei.

\bsk
\baselineskip = 12pt
\text{\ni 2. BACKGROUND}
\ssk
\ni     

	In 1928 Robert Millikan proclaimed that cosmic rays are neutral quanta,
the ``birth cry" of atoms created by elemental synthesis from primordial
hydrogen spread throughout the universe.  Seventy years later, at the
recent 1998 APS meeting in Columbus, Srinivas Kulkarni presented a lecture
entitled ``Gamma Ray Bursters: Dying Cries  from the Distant Universe".
This remarkably glorious era for high energy astrophysics all began with
little fanfare on 7 August 1912 when a balloon-borne gondola ascended from
a field near Aussig, Austria; in the gondola were the physicist Victor Hess,
two helpers and three electroscopes.  As expected, for the first kilometer
the electroscope discharge rate decreased somewhat with altitude as the
gondola moved away from the presumed traces of radioactivity in the earth's
crust.  However, at higher altitudes the trend reversed; at an altitude of
about 5 kilometers the rate was four times faster than on the ground.
Quite correctly, Hess concluded that this arose from an ionizing flux
coming from above and conjectured that it was ultimately due to penetrating
radiation falling upon Earth from somewhere beyond the atmosphere.  During
the next three decades, however, the exact nature of the primary radiation
remained a mystery [for fascinating accounts of this early history see
``Cosmic Rays" by Rossi (1964) and Friedlander (1989)].  First of all, was
this radiation neutral (e.g., ``ultra" gamma-rays) or charged?  For a
substantial time there were two schools on this.  As already noted,
Millikan (who invented the name ``cosmic rays") favored the interpretation
that they are neutral.  Arthur Compton, his collaborators and eventually
other researchers (including Millikan) concluded that the radiation was
positively charged, as deduced from using the earth's dipole field as a
magnetic analyzer.  Today we know that the cosmic ray flux also involves
ionizing components of negative charges (electrons) and photons (gamma rays
and X-rays) as well as nuclei; Table 1 gives the various components, listed
in order of energy flux impinging upon the atmosphere.

	In order to provide some historical perspective on the opening of new
observational windows, it is interesting to consider some of those modern
cosmic discoveries readily categorized as ``expected" or ``unexpected"(see
Table 2).  Although we anticipated neither gamma ray bursts nor
extragalactic blazar gamma sources, we definitely did expect diffuse
galactic gamma rays.  After all, gamma radiation is a necessary consequence
of high energy cosmic ray protons ($>0.3$ GeV) interacting with interstellar
matter; the resultant neutral pion products yield a flux of photons $>100$
MeV well traced over the entire Milky Way via the EGRET instrument on CGRO
(Fichtel et al. 1993; Hunter et al. 1997).  In sharp contrast, X-ray
astronomy came as a complete surprise.  We knew of the interstellar medium
(ISM) but did not expect to encounter a local hot X-radiating interstellar
plasma, now known from ROSAT mapping to be a lasting imprint of past
explosive events in our region of the Milky Way (Snowden et al., 1997).
Who would have thought that the first direct evidence for the shock
acceleration of relativistic cosmic rays would come via X-ray astronomy?
Yet, ASCA X-ray data on the supernova remnant SN1006 did just that (Koyama
et al., 1995).  It's becoming very clear that the charged particle and
electromagnetic components of cosmic radiation are inexorably intertwined;
as emphasized in this presentation, complementary studies of them can be
very revealing.
\begin{table}
\caption{Table 1. Cosmic Radiation: Principal Ionizing Components (energy flux order)}
\begin{center}
\begin{tabular}{lll}
\hline\hline

1 &	Nucleonic (galactic)			$E >20$ MeV/nucleon  \\
2.&	Electron (galactic),			relativistic   \\
3.&	X-rays (extragalactic)    \\
4.&	X-rays (soft galactic)    \\
5.&	Gamma-rays (diffuse galactic)     \\
  &		\ \ $\bullet$	$h\nu >100$ MeV $\rightarrow$ $E > 0.3$ GeV/nucleon  \\
  &	 \ \ $\bullet$		$h\nu <100$ MeV $\rightarrow$ electron bremsstrahlung  \\
  & \ \ $\bullet$ spectral lines (nuclear, $e^+e^-$ ) \\
6.& 	Gamma-rays (extragalactic)    \\
7.& 	X-rays (galactic binaries, SN...)   \\
8.& 	 X-rays (diffuse non-thermal galactic):   \\
  &	\ \ $\bullet$  subrelativistic cosmic ray bremsstrahlung?     \\
  &	\ 	\ \ \  	$h\nu \ge$ 10 keV$\rightarrow E\ge$ 20 MeV/nucleon \\
  &	\ \ $\bullet$	atomic K lines \\

9.&	Cosmic rays (extragalactic) 		$E >10$ Joules/each   \\
\hline\hline
\end{tabular}
\end{center}
\end{table}

\begin{table}
\caption{Table 2. Historical Perspective: New Windows on the Universe: Cosmic Discoveries}
\begin{center}
\begin{tabular}{ll}
\hline\hline

EXPECTED  &	UNEXPECTED \\
\hline 
	An interstellar medium	& 	A hot x-radiating plasma\\
	Gamma-ray Astronomy	&	Gamma-ray bursts\\
	Black holes		&		X-ray Astronomy\\
	Neutron Stars		&		Pulsars\\
	Gravitational radiation		&		Binary pulsar\\
	Gravitational lensing	&		Dark matter\\
	The Milky Way		&	Galaxy clustering\\
				&		Supermassive Black Hole Galactic Nucleus\\
	Evolution		&		Radio Astronomy\\
				&		Quasars\\
				&		Cosmic X-ray Background\\
	``Big Bang" relic Cosmic	&\\	
	Microwave Background 	&		A distortionless black body spectrum\\
	Primordial gas in clusters	&	Iron K-line emission in clusters\\
	Baryon Symmetry	&		Matter, matter everywhere\\
	Extragalactic Cosmic Rays&		Too energetic\\
\hline\hline
\end{tabular}
\end{center}
\end{table}

	Astronomy is replete with examples in which the most significant advances
or the most astounding discoveries arose with the opening of new
observational windows, partly by design and partly by chance.  We cite two
recent examples from high energy astrophysics; see Table 2 for others.
RXTE was designed to have the high throughput and timing capability needed
to measure rapid variability in low-mass X-ray binaries, but it was the
unexpected millisecond oscillations observed that opened up a fundamental
new phase in X-ray astronomy, one that probes neutron stars in the
dynamical and gravitational regime needed to constrain the equation of
state of neutron-star matter and to measure the neutron star mass in the
strong-field limit of general relativity (Kaaret, Ford \& Chen, 1997; Zhang,
Strohmeyer \& Swank, 1997).  Clearly, the particular complement of
synergetic instruments aboard BeppoSAX is the main factor that has made it
possible to exploit the unexpected X-ray afterglow of gamma-ray bursts for
providing sufficiently prompt accurate positions, those needed for the
precise optical measurements that are establishing their cosmological
origins (Costa et al.,1997; Metzger et al.,1997).

\begin{table}
\caption{Table 3. Space  Physics  Chronology:  1958 -}
\begin{center}
\begin{tabular}{lll}
\hline\hline
	1958	&	Van Allen radiation belts 		&space hazard\\
	1959	&	Solar wind				&cosmic ray barrier\\
	1960	&	Cosmic ray electrons			&spectral neutrality\\
	1962	&	Cosmic X-rays			&accretion, black holes\\
	1963	&	Quasars				&compact galactic nuclei\\
	1964	&	Subrelativistic cosmic rays	&	ISM ionization \& heating\\
	1965	&	Microwave background		&early universe\\
	1968	&	Pulsars 			&	neutron stars\\
		&		Solar neutrinos		&	nuclear furnace\\
		&		Hot ISM			&	soft X-radiation\\
	1972  ('61)&	Cosmic gamma-rays		&	galactic cosmic ray tracer\\
	1973  ('68)&	Gamma-ray  bursts		&	new physics /astrophysics?\\
	1973	&	Fe X-ray K lines 		&	hot plasma spectroscopy\\
	1974	&	Binary pulsar  			&  gravitational radiation\\
	1978 ('33)&	Missing mass			&	galactic, clusters\\
	1984	&	$^{26}$Al gamma-ray line		&	nucleosynthesis tracer\\
	1985	&	Gravitational lensing		&	dark matter\\
	1987	&	Non-solar neutrinos		&	SN1987A\\
	1991	&	Extragalactic cosmic rays 	& 	new physics/astrophysics?\\
	1992	&	Blazar gamma-rays		&	beamed emission\\
	1993	&	Machos				& massive compact halo  objects\\
	1994    &       Microquasars                    & relativistic stellar jets \\
	1995	&	Cosmic ray  accelerator		& SN1006\\
	1996&		Helioseismology			& solar inner structure\\
	                 &	HST deep field	& early star/galaxy history  \\
	1997		&Gamma burst afterglow		& cosmological origin\\
			&	Millisecond QPOs	&general relativity strong field limit\\
\hline\hline
\end{tabular}
\end{center}
\end{table}

	Astrophysical studies from space platforms began with the Explorer 1
discovery of the Van Allen radiation belts in 1958 (Van Allen, Ludwig, Ray
\& McIlwain, 1958); Table 3 lists some of the most significant space physics
milestones since then.  Although the advent of orbiting observatories
brought us above the atmospheric barrier that had previously hampered the
study of primary cosmic rays, this discovery made us aware of the radiation
hazard these belts posed for the sensitive detectors involved and made us
consider how to minimize such encounters.  The 1959 detection of the solar
wind plasma with Lunik I \& II (Shklovskii, Moroz \& Kurt, 1960) and studies
of solar driven cosmic ray modulation effects of the wind
(Balasubrahmanyan, Boldt \& Palmeira, 1965, 1967) made us realize that
getting above the atmosphere was not sufficient and that, eventually, this
additional formidable barrier would have to be surmounted for the proper
study of the interstellar cosmic ray flux.  The first clear measurements of
the low-energy spectum of cosmic rays, those arriving here at 1 AU,
involved observations made far from Earth, albeit well within the
heliosphere (McDonald \& Ludwig, 1964).  So far, deep space probes have yet
to reach bonafide interstellar space ($>120$ AU), beyond the reach of solar
wind effects (Axford, 1992).  By the year 2000 the distances observed with
Voyagers 1 \& 2 will be 76 and 61 AU respectively.  However, the solar wind
has to this day still prevented us from measuring flux values for low
rigidity cosmic rays at all close to being as large as those values
expected for the corresponding interstellar particles outside the
heliosphere (McDonald, 1998).  After reaching beyond the heliosphere, the
ultimate remaining barriers in directly detecting an all-inclusive sample
of subrelativistic cosmic ray nuclei are {\em low energy} and {\em short lifetime}.
The average high energy cosmic ray traverses $\sim $10 g/cm$^2$ of ISM matter in
$\sim 10^7$ years before being lost (e.g., by escape from galactic confinement).
Hence, for a cosmic ray particle to reach our neighborhood of the Milky Way
from its place of origin there must be sufficient energy to traverse a
minimal ISM columnar density of matter and, if a radioactive nucleus,
sufficient lifetime ($\ge 10^7$ years) before decay.  As we shall see in the next
section (Section 3) the emission of characteristic X and gamma rays in basic
radiative processes is crucial in providing us with the means needed to
achieve remote sensing of otherwise inaccessible cosmic rays and thereby
obtain a significantly more comprehensive sample.

\bsk
\baselineskip = 12pt
\text{\ni 3. CONNECTIONS}
\ssk
\ni     

	The lowest energy cosmic rays in the Milky Way are to be found among those
fresh products of nucleosynthesis that have propagated into the
interstellar medium but have not yet become thermalized.  The
quintessential example is the $\beta$-radioactive nucleus $^{26}$Al first observed
with the solid-state (Ge) spectrometer on HEAO-3 (Mahoney, Ling, Wheaton \&
Jacobson, 1984) by detection of the 1.8 MeV gamma ray spectral line from an
excited state of the daughter  $^{26}$Mg.  Having a relatively short average
decay lifetime ($1.0 \times 10^6$ years), an order of magnitude less than the
confinement lifetime of energetic cosmic rays, and suprathermal velocities
of only $\sim 500$ km/s (Naya {\em et al.}, 1996), 
the distribution of such $^{26}$Al provides a
current snapshot of nucleosynthesis sites throughout the galaxy, as
obtained by imaging of 1.8 MeV gamma rays with the COMPTEL instrument on
CGRO (Diehl et al. 1995).  The next generation of gamma-ray spectrometers,
such as the INTEGRAL spectrometer (von Ballmoos, 1995), should give the
improved maps needed to more completely trace the current nucleosynthesis
activity in the Milky Way, especially for regions of relatively low surface
brightness.  This nucleosynthesis of $^{26}$Al in stars and stellar explosions
is to be distinguished from that produced at much higher energies as fully
stripped spallation products from energetic cosmic ray transport in the
ISM; Simpson and Connell (1998) have suggested searching the galactic halo
for a highly broadened 1.8 MeV line from such transrelativistic nuclei.  In
sharp contrast, it's important to note that a subrelativistic $^{26}$Al ion
injected into the ISM rapidly reaches a charge equilibrium, determined
solely by its velocity, via the competing processes of electron capture and
loss (Pierce \& Blann, 1968).  For aluminum at 500 km/s ($\sim 1$ keV/nucleon) the
average effective charge of the ion is thereby only $\sim$0.5 (i.e.,
essentially a neutral atom).  In this situation we must consider the
alternate decay mode of $^{26}$Al to $^{26}$Mg via electron capture (rather than
positron emission), one which then occurs 18\% of the time.  The attendant
removal of a K shell electron leads to a 1.25 keV K$\alpha$ X-ray from the atomic
transition that replaces the missing K shell electron for the Mg daughter.
This narrow X-ray spectral line is a complementary tracer of
subrelativistic $^{26}$Al in the local Milky Way (within $\sim$ 3 kpc), one that has
the potential of providing the high resolution spatial/spectral
measurements needed to determine the detailed distribution, velocity,
ionization state and chemical setting of these freshly synthesized nuclei.
In particular, it has been suggested (Lingenfelter, Ramaty \&
Kozlovsky,1998) that the subrelativistic $^{26}$Al detected actually resides in
rapidly moving grains of Al$_2$O$_3$.  We should note that unit optical depth for
the photo-absorption of a 1.25 keV X-ray in Al$_2$O$_3$ would be 1.6 $\mu$m, not very
much more than the largest sizes of usual grains; hence, this presents us
with the possibility of using the x-ray signal for some direct chemical
diagnostics.

	Preliminary results from COMPTEL observations of some galactic regions suggesting 
an enhancement in 3 - 7 MeV gamma-ray emission (Bloemen, 1994) have led
to the realization that such a signature could indicate the presence of
much freshly synthesized $^{12}$C and $^{16}$O undergoing inelastic
collisions with ambient H and He with sufficient energy ($\ge 20$
MeV/nucleon) to excite nuclear levels that decay via characteristic
gamma line emission, substantially broadened (Kozlovsky, Ramaty \&
Lingenfelter,1997).  In this scenario the 0.5 - 1 keV X-ray emission
from these regions could by dominated by somewhat broadened atomic
K$\alpha$ lines associated with electron capture by fast C and O ions
at $\beta \equiv v/c \le 0.1$, mainly the 0.57 keV line from O$_{VII}$
and the 0.65 keV line from O$_{VIII}$ (Pravdo \& Boldt, 1975;
Tatischeff, Ramaty \& Kozlovsky, 1998); sufficiently sensitive searches
for this emission are yet to be made.  Surveying the Milky Way for such
X-ray lines would trace the location in our galactic neighborhood ($<
1$ kpc) of possible regions of enhanced subrelativistic cosmic
rays.

	What is the evidence for a significant interstellar population of
subrelativistic ($\beta <<1$) cosmic ray nuclei broadly distributed throughout the
galactic disk?  Relative to the local interstellar flux of such nuclei
beyond the heliosphere, the small flux arriving here at $\sim$ 1 AU corresponds to
severe attenuation by the solar wind.  For a range of sufficiently low
rigidities ($R$) the attenuation factor ($F$) may be approximated by

\begin{equation}
			F \approx exp(-K/\beta),
\end{equation}
where $K \ge 1$, depending mainly on the overall spatial extent of the wind,
its velocity and frozen-in magnetic field structure, but only weakly on R
(Parker, 1963).  The upturn in the spectrum of subrelativistic nuclei
observed below $\sim$ 20 MeV/nucleon arises from an ``anomalous" component now
recognized as being energized within the heliosphere (Kinsey, 1970; Fisk,
Kozlovsky \& Ramaty, 1974; Jokippi \& McDonald, 1995), thereby masking the
identification of any subrelativistic interstellar cosmic rays with $\beta < 0.2$
that might arrive here at 1 AU.  As here emphasized next, we can gain
access to the illusive interstellar spectrum of subrelativistic cosmic rays
by appropriate X-ray measurements that, in effect, provide the direct
remote sensing of these particles called for.

	Recent RXTE observations (Valinia \& Marshall, 1998) of the apparently
diffuse X-radiation (10$\rightarrow$ 60 keV) from the galactic ridge [first observed in
the 2 $\rightarrow$ 10 keV band by Bleach et al. (1972)] now indicate a non-thermal disk
component having a power-law spectrum
\begin{equation}
			dL_x/d(h\nu) \propto (h\nu)^{-1.3}
\end{equation}

with a luminosity
\begin{equation}
			L_x (10-60 \ keV) =1.5 \times 10^{38} \ \rm erg/s.
\end{equation}
Is this luminosity providing us with a measure of the input required to
produce the observed ionization of the ISM?  To evaluate this we note that
the intensity of the galactic H$\alpha$ background measured at high galactic
latitudes implies an average hydrogen recombination rate of $4 \times 10^6 s^{-1}$ per
cm$^2$ of the galactic disk (Reynolds, 1984).  With the required dissipation
of $\sim$40 eV per electron-ion pair generated, the corresponding total galactic
ionizing input ($Q$) implied is

\begin{equation}
			Q = 7 \times 10^{41} \ \rm ergs/s.
\end{equation}

Taking $L_x$ as the appropriate radiative measure of this total required rate
of energy dissipation (i.e., the power input to produce the observed
ionization) allows us to  evaluate the associated radiative yield ($Y$), viz:
\begin{equation}
			Y \equiv L_x/Q = 2 \times 10^{-4}.
\end{equation}

That the value of this radiative yield is comparable to what is expected
for an energetic ionizing particle in the ISM with $\beta \approx 0.2 - 0.5$, suggests
that the observed 10-60 keV disk radiation arises mainly via the process of
suprathermal proton (\& alpha) bremsstrahlung for a relatively high flux of
cosmic ray particles at 20-120 MeV/nucleon (Boldt \& Serlemitsos, 1969).
The observed X-ray spectrum (eq. 2) then implies an interstellar cosmic ray
spectral form for the particle flux given by
\begin{equation}
\delta J/\delta E \propto  E^{-1.3} \ \ \ \ \ \ \ \ ({\rm for}\  E = 20-120\  {\rm MeV/nucleon}).
\end{equation}
	For the particular band considered, this spectrum (eq. 6) is a remarkably
good approximation to the similarly restricted portion of the interstellar
cosmic ray spectrum proposed by Balasubrahmanyan et al.(1968).  That
spectral model corresponds to cosmic ray propagation through the ISM with
an exponential distribution of path lengths having a mean value of 6 g/cm$^2$
for injected particles whose spectra at the input sources exhibit a single
power law that extrapolates from what is directly observed at $E >> 1$
GeV/nucleon  all the way down to $E < 100$ MeV/nucleon.  As shown in that
paper, comparing this interstellar spectrum with that observed, at solar
minimum, implies that the modulation parameter for eq. 1 is $K=2.7$.
Integrating the interstellar cosmic ray spectrum then yields a value for
$Q(>20$ MeV/nucleon) that is a  substantial fraction of that required (eq.4)
and an energy density of $\sim$3 eV/cm$^3$, higher than previously considered
(Gloeckler \& Jokippi, 1967).  If this conclusion is valid, then cosmic rays
play a major (possibly dominant) role in the dynamics of the ISM.  An
alternate possibility that the power-law source spectra extrapolate to low
energies on the basis of total energy (rather than kinetic energy) would
yield a cosmic ray flux whose energy density is only 0.6 eV/cm$^3$; however,
the interstellar spectrum would then be incompatible with the observed
X-ray spectrum and correspond to a $Q$ value two orders of magnitude too
small.  Confirmation of the RXTE result and this interpretation of it is
clearly essential.  The increased spatial resolution and bandwidth of the
INTEGRAL spectrometer should provide the information needed to establish
that the hard X-ray ridge emission is indeed mostly diffuse non-thermal radiation
and give us precise maps of its surface brightness distribution that we can
use to trace subrelativistic cosmic rays throughout the ISM of the Milky
Way. Galactic ridge gamma-ray lines from  excited states of interstellar $^{12}$C and
$^{16}$O, those induced via inelastic collisions by cosmic ray particles, should
further define the subrelativistic nucleonic component (Pohl, 1998).
Although the volume emissivity of the galactic ridge in X-rays is
more than an order of magnitude greater than that expected from the
bremsstrahlung of cosmic ray electrons in the ISM, relativistic electrons
could well account for most of the galactic bremsstrahlung emission in the
1-100 MeV gamma ray band (Strong et al. 1994).

\begin{table}
\caption{Table 4.  Energy Density [u (eV/cm$^3$)] of Local Fields (outside heliosphere)}
\begin{center}
\begin{tabular}{lll}
 \hline\hline

 1 &  Cosmic rays ($>20$ MeV/nucleon)	& $\ge	1$    \\
 2 &  Galactic magnetic field ($B^2/8\pi$)&       $\sim 1$ \\
 3 &  Starlight (galactic)		&	$0.3$    \\
 4 &  2.7$^\circ$ K microwave background	&$	0.3$    \\
 5 &  1.9$^\circ$ K neutrino background 	&	$0.1$  \\
 6 &  Cosmic ray electrons 		&	$\le 0.1$  \\
 7 &  Extragalactic IR background  ( $\lambda >140 \ \mu m$)&	$4\times 10^{-3}$  \\
 8 &  Extragalactic objects  ($\lambda =0.36-2.2\ \mu m$)&	$3\times 10^{-3}$  \\
 9 &  Gravitational radiation background    [$\nu (u_\nu)$]	  &	$\le 6\times 10^{-4}$  \\
 10 &  Quasar light  (inferred bolometric)  	&	$3\times 10^{-4}$  \\
 11 &  Cosmic (extragalactic) X-rays ($>1$ keV) &	$6\times 10^{-5}$  \\
 12 &  Soft galactic X-rays  ($<1$ keV) 	&       $1\times 10^{-5}$  \\
 13 &  Extragalactic (blazar) gamma-rays 	&	$9\times 10^{-6}$  \\
 14 &  Galactic gamma rays  ($>100$ MeV)  	&	$8\times 10^{-6}$  \\
 15 &  Galactic X-rays ($>1$ keV)  	&	$6\times 10^{-6}  (2\times 10^{-7}$ unresolved) \\
 16 &  Extragalactic cosmic rays  ($> 10^{20}$ eV)&       $\sim 10^{-9}$ \\
 \hline\hline
    &   & \\
    &   & \\
 References & & \\
 \end{tabular}
 \begin{tabular}{ll}
        5. & 	Gravitation and Cosmology (Weinberg , S. ,1972 , John Wiley \& Sons,
New York)  \\
      & p. 528, 537  \\
	6.&	Interstellar cosmic ray electron flux estimated from Fig. 8 in Strong
et al. (1994) \\
	7.&  From COBE  results obtained with DIRBE (Hauser, M. et al., 1998, ApJ, in
press) and  \\
	  & FIRAS (Fixsen, D. et al., 1998, ApJ, in press) \\
	8.&	Extragalactic background light estimated from HST deep field resolved
sources and \\
	  & complementary ground based data (Pozzetti, L. et al. 1998,
MNRAS, in press; \\
    & Hauser, M. 1998,\\
   &  personal communication)\\
	9.&	Kaspi, V., Taylor, J. \& Ryba, M., 1994, ApJ, 428, 713 \\
	10.&	From the  bolometric luminosity density of quasars suitably integrated
over all  \\
	  &  redshifts (Chokshi \& Turner, 1992) \\
	11.&	Boldt (1987) \\
	12.& 	Obtained from ROSAT all-sky surface brightness data (Snowden, 1998) \\
	13.&	Sreekumar et al. (1998) \\
14.&	Hunter et al. (1997) \\
15.&	Boldt (1974) \\
16.&	Cronin (1997) \\
\end{tabular}
\end{center}
\end{table}

\bsk
\baselineskip = 12pt
\text{\ni 4. FULL CIRCLE}
\ssk
\ni     

In tabulating the estimated energy density in every radiation field of
extrasolar origin that obtains here in our region of the Milky Way (outside the
heliosphere) we note (see Table 4) that the galactic cosmic ray nucleonic
component is the largest of all.  The opposite extreme (at the bottom of
the list), that of ultra-high energy extragalactic cosmic rays  ($>10^{20}$ eV
each), is manifested right here in the atmosphere of Earth at a rate
somewhat less than one per km$^2$ per decade.  This time using the atmosphere
to good advantage (i.e., as a detection  medium), Bird et al. (1995) have
observed the atmospheric fluorescence produced by an extensive air shower
initiated by a $3.2 \times 10^{20}$ eV ($\sim$50 Joule) cosmic hadron, the highest energy
particle yet detected in nature.  The Larmor radius for this energetic a
proton in the galactic magnetic field would be $\sim 150$ kpc; in the
intergalactic field it would be $\ge 300$ Mpc.  Hence, such a particle must be
extragalactic in origin.  Furthermore, for a source distance much less than
300 Mpc its observed vector velocity would still be close to the initial
direction of emission.  However, due to collision losses with the 2.7$^\circ$ K
background radiation field (Greisen, 1966; Zatsepen \& Kuz'min, 1966), a
proton this energetic could not have survived from a source more distant
than 50 Mpc (Elbert \& Sommers, 1995).  And there are no suitable active
galaxies in the right direction that are close enough to be viable source
candidates.  We emphasize here, however, that the present electromagnetic
radiation field arising from accretion-driven AGNs in the past implies the
existence of a substantial ``local" population of apparently dormant
galactic nuclei that harbor spinning supermassive black holes.  These could
be the sites of hidden dynamos sufficient for powering the required
production of those cosmic rays whose energy is presently regarded as
astrophysically excessive (Boldt \& Ghosh 1998).  Here, in this very extreme
instance of high energy astrophysics, we again see the involvement of a
profound cosmic ray/astronomy connection and the role of photons in
revealing the origins of puzzling new exotic phenomena.  We seem to have
come full circle with this adventure.  Our earth's atmosphere, which
initially was a seemingly insurmountable barrier for us in the detection of
primary cosmic rays, is now the most effective detection medium available
for the very highest energy quanta in nature.  Although it would appear
that the most energetic particles observed here must be produced in the
present epoch of the universe, we look to the overall cosmic background of
accretion-driven electromagnetic radiation for providing the integral
measure, over all past epochs, that defines the local candidate source
population of ``hidden" supermassive black holes.

	Even though local dormant quasar remnants are manifestly under-luminous,
their underlying supermassive black holes are likely to be sufficiently
spun-up [i.e., after their many  ``Salpeter" time units of accretion history
(Thorne 1974; Rees 1997)] to possibly serve as high-energy accelerators of
individual particles.  In this scenario (cf., Blandford \& Znajek 1977)
externally produced magnetic field lines threading the event horizon of
such black holes would, by virtue of the induced rotation, generate an
effective electromotive force characterized by: $emf  \propto c B R$, where $B$ is
the magnetic field strength and $R$  is the effective range over which the
concomitant electric field is applicable.  Scaling to the magnitude for
this impressed $B$  field considered by Macdonald and Thorne (1982) and
taking $R \approx R_g (\equiv GM/c^2)$, the gravitational radius, the expected value for
the $emf$  is then here estimated as
\begin{equation}
		emf  \approx 4 \times 10^{20} B_4 M_9 \  {\rm volts,}
\end{equation}
where  $B_4 \equiv B/(10^4$ Gauss) and $M_9 \equiv M/(10^9M_\odot$).  We note that radiative
losses for electrons in such a dynamo greatly exceed those for protons. 
 Hence, radiative cascades (basic to the Blandford -
Znajek mechanism), such as attend the process of electron acceleration,
would not constitute a comparable limiting factor in the present scenario
for the acceleration of the relatively few (favorably disposed) protons
that need to achieve an energy close to that of exploiting the full
voltage.  The situation might well be one in which the accelerator is not
operational in the mode in which quasi-steady conversion of the hole's
rotational energy into that of luminous radio jets is possible, but where
acceleration of individual protons by it may occur, perhaps sporadically
(Boldt \& Ghosh 1998).

	What are the environmental circumstances of the black hole nuclei in dormant 
	quasar remnants and are they conducive to having sufficient accretion for
sustaining the magnetic fields needed for this generator?  The unusually
low luminosities associated with the supermassive black holes at the
centers of nearby bright elliptical galaxies (Loewenstein et al. 1998;
Fabian \& Rees 1995; Fabian \& Canizares 1988) have been explained in terms
of radiatively inefficient advection-dominated flow (Mahadevan 1997;
Naranan 1997), even when there is ample ambient gas available for
accretion.  A recent critical reassessment  (Ghosh \& Abramowicz 1977) of
the likely strengths of magnetic fields threading the horizons of
accretion-disk fed black holes leads to the conclusion that these strengths
are somewhat lower than previously considered; considerations of
advection-dominated disks as opposed to standard disks could further lower
the estimated field strength.  Taking $B_4<1$  in eq. 6 then implies that
achieving the desired $emf$ would require $M_9 >1$. What is the evidence for
such supermassive black holes in a substantial present-epoch population of
dormant galactic nuclei?  In the nearby universe, there is a drastic
paucity of quasars such as the extremely luminous ones ($L \ge 10^{47}$ ergs/s)
evident at large redshifts ($z > 1$), those with putative black hole nuclei
having masses $\ge 10^9 M_\odot$.  Nevertheless, the local number of dead quasars
associated with the same parent population (Schmidt 1978; Small \&
Blandford, 1992; Richstone et al., 1998) is expected to be relatively
large.  And now there is also direct stellar-kinematic evidence for
individually identified massive dark objects (MDOs) at the centers of
several nearby inactive galaxies (Kormendy \& Richstone, 1992; Magorrian et
al. 1998).

	As emphasized by Soltan (1982) for a standard Friedmann cosmology, the
total background radiation arising from the entire history of accretion fed
AGN emission gives us a direct integral measure of the total mass built-up,
that which is now present locally in the form of black holes in AGNs and
their remnants.  Independent of the Hubble constant ($H_\circ$) and deceleration
parameter ($q_\circ$), the total mass density built-up by this accretion over all
cosmic time (t) is given by
$$
	\rho ({\rm growth}) ={ 2.6 \times 10^7 (\epsilon^{-1} - 1) (1+ \langle z \rangle ) [u(eV/cm^3)] } \ M_\odot/(Mpc)^3 \eqno(8a)
$$
where:
$$
\epsilon({\rm radiation\  efficiency})  \equiv  L /[c^2(\delta M/\delta t)_{\rm accreted}] \eqno(8b) $$ 
$$\  \ \ u({\rm energy\  density})  \equiv  4\pi I /c = \int [(n\langle L \rangle )/(1+z)] \delta t, \eqno(8c) $$ 
$4\pi I$ is the omnidirectional bolometric background flux arising from AGNs,
and (n$\langle L \rangle$) is the comoving bolometric luminosity density.

	 Studies of the X-ray sky indicate a pronounced extragalactic cosmic X-ray
background (CXB) that arises mainly from accretion powered AGN emission at
previous epochs (Boldt 1987; Fabian \& Barcons 1992).  By correlating
surface brightness fluctuations of the CXB with IRAS galaxies Barcons et
al. (1995) find that the present-epoch 2-10 keV luminosity density is
dominated by Seyfert 1 galaxies ($L_x >10^{42}$ ergs/s).  Padovani, Burg and
Edelson (1990) have determined that the local mass density in the form of
Seyfert 1 nuclei is $\sim 6 \times 10^2 M_\odot/(Mpc)^3$, half of which arises from black holes
of mass $M > 3 \times 10^7 M_\odot$ and essentially none from any possible AGN black holes
of mass $M > 2 \times 10^8 M_\odot$.  For an accretion powered X-radiation efficiency $\le$10\%
( i.e., $\epsilon \le  0.1$ in eq. 8a) the observed CXB energy flux implies the
build-up of a local mass density $> 14 \times 10^3(1+\langle z \rangle ) M_\odot /(Mpc)^3$,
where $\langle z \rangle \ge 1$.
This density is clearly much larger than that for Seyfert 1 nuclei.  In
order to account for the flux and spectrum of the CXB it is necessary to
invoke a supplementary source population that somehow makes a substantial
redshifted contribution to the observed CXB without making much of a
contribution to the local 2-10 keV luminosity density (Boldt \& Leiter 1995;
Boyle et al. 1998).  This needed additional component (over and above that
from Seyfert 1 nuclei) could arise from a population not at all represented
locally [e.g., precursor AGNs necessarily at large z (Boldt \& Leiter 1995)]
and/or Seyfert 2 nuclei whose emission below $\sim$10 keV is strongly attenuated
by absorption (Madau, Ghuisellini \& Fabian 1994; Comastri et al. 1995).
The unified Seyfert AGN model (Madau, Ghuisellini \& Fabian 1994) would
imply that the present-epoch mass spectrum for Seyfert 2 nuclei would be
the same as for Seyfert 1, renormalized by the ratio of their local number
densities $(n_2/n_1)$.  In particular, for these scenarios, using $(n_2/n_1) \le 4$
(Madau, Ghuisellini \& Fabian 1994; Comastri et al. 1995) implies a total
local Seyfert mass density $\le 3 \times 10^3 M_\odot (Mpc)^3$, much less than that implied
by the CXB flux.  In these models, the bulk of the local mass density
related to the CXB would be accounted for by dormant Seyfert remnants
having the same average mass ($\sim 2 \times 10^7 M_\odot$) as the AGNs.  Clearly, the black
holes associated with Seyfert remnants are not massive enough to be viable
candidates for the high energy dynamos we are looking for.  The black holes
associated with blazar remnants are likely to be appreciably more massive.
If blazar emission is powered by accretion, then we can use eq. 8a, with
the gamma-ray background energy density (Table 4), to evaluate that the
local mass density in the form of associated black holes is $\ge 2 \times 10^3(1+ \langle z \rangle)
M_\odot/(Mpc)^3$, where 
$\langle z \rangle \ge 1$ and $ \epsilon \le  0.1$; this blazar remnant density is
somewhat larger than that associated with active (Seyfert) galaxies.

	Considering a radiative efficiency of 10\% for the accretion powered
bolometric luminosity of quasars, Chokshi and Turner (1992) have calculated
the mass built up over all earlier epochs and thereby estimated that the
expected local mass density in compact galactic black hole nuclei is two
orders of magnitude greater than that accounted for by Seyfert galaxies.
They conclude that over 10\% of this density is associated with black holes
of mass $ >6 \times 10^8h^{-2}M_\odot$, where $h \equiv H_\circ / [100 km \ s^{-1}(Mpc)^{-1}].$
As emphasized by
Chokshi and Turner (1992), the local universe is expected to be well
populated by currently inactive remnants of quasars.  Based on the mass
function described by them we have estimated that, for $h \approx 0.5$, there
should be about a dozen or more quasar remnant black holes of mass $>10^9M_\odot$
within 50 Mpc.  These quasar remnant expectations are consistent with being
lower limits to the number of corresponding supermassive black holes
inferred from a recent comprehensive study of massive dark objects (MDOs)
at the centers of 32 nearby galaxy bulges (Magorrian et al. 1998).  In this
connection we note that the number of MDOs within 50 Mpc identified in
their sample as being more massive than $10^9M_\odot$ is already 8, comparable to
the total number of Seyfert 1 AGNs out to that distance. This is a lower
limit to the total number of such supermassive MDOs within this volume
since their sample of MDOs at the centers of nearby galaxy bulges is
incomplete, albeit sufficiently large for the correlations sought by them
(Magorrian 1998).  Using the luminosity function for field galaxies
(Efstathiou, Ellis \& Peterson, 1988) to estimate the incompleteness of
their sample suggests that the corrected number of supermassive MDOs could
well be an order of magnitude greater than that so far observed.  It is
interesting to note that in their sample of 32 nearby MDOs (Magorrian et
al. 1998) five are associated with compact objects somewhat more massive
than $10^{10}M_\odot$.

	The sample of about 100 extraordinary cosmic ray events with energy $\ge
2 \times 10^{20}$ eV expected with the upcoming Auger extensive air-shower array
(Cronin 1997) will come from nearby sources and, if protons, will point
accurately to the directions of origin (i.e., owing to the correspondingly
large particle Larmor radius in the weak intergalactic magnetic field).
Candidate galaxies within the acceptable pixels would then be searched for
stellar-dynamical evidence for central supermassive black hole nuclei [such
as the MDOs discussed by Kormendy \& Richstone (1995), Kormendy et al.
(1997) and Magorrian et al.(1998)], here taken to be indicative of the dead
quasar sources of the highest energy cosmic rays.  If such a correlation is
clearly established, and the lack of correlation with strong radio sources
persists, it would imply that the existence of a black hole dynamo is not a
sufficient condition for the presence of pronounced jets.  The OWL
(Orbiting Wide-angle Light-collectors) space borne NASA mission planned for
observing, from above, those air showers induced by the highest energy
cosmic rays is anticipated to have the sensitivity for accumulating an
order of magnitude more such events than expected with the Auger array
(Streitmatter 1998).  If in fact no real correlation is found with
sufficiently nearby MDOs, one would then have to pursue more exotic
particle physics possibilities, disregarded in this present discussion, in
which: 1) the primary hadronic particles are produced at ultra-high
energies in the first instance, typically by quantum decay of some
supermassive elementary particles related to grand unified theories (Sigl
et al. 1995; Kuz'min and Tkachev 1998) or  2) a new neutral massive ``$S^\circ$"
hadron ($m_\circ c^2 > 2$ GeV) is produced whose energy loss to the cosmic microwave
background is relatively small, even for very remote sources (e.g.,
quasars) at cosmological distances (Chung, Farrar \& Kolb, 1998; Farrar \&
Biermann, 1998).  If one of these new particle scenarios turns out to be
the case, then cosmic ray research will have come full circle, back again
to its role of providing the leading thrust towards new fundamental physics
as well as astrophysics. ``When we want to learn more about nature,
exploration of extremes has consistently added to our knowledge and brought
us surprises (Cronin, 1997)"

\bsk
\baselineskip = 12pt
{\abstract \ni ACKNOWLEDGMENTS
I thank G. Palumbo for  suggesting this historical overview of cosmic ray
astrophysics and his confidence in my doing it.
In preparing this presentation I have had the benefit of valuable inputs
from D. Bertsch, J. Cronin, M. Hauser, S. Hunter, K. Jahoda, P. Kurcznski,
D. Leiter,
G. Ludwig, J. Naya, J. Magorrian, F. McDonald, M. Persic, M. Shapiro, S.
Snowden, T. von Rosenvinge, A. Valinia and W. Zhang.  I am especially
grateful for particularly illuminating discussions over many prior years
with V. K. Balasubrahmanyan concerning our early studies of subrelativistic
cosmic rays which, as exhibited in this lecture, still constitute a
challenge of central interest to me.  }

\bsk
\baselineskip = 12pt


{\references \ni REFERENCES
\ssk

\ref Axford, W. I. 1992, in The Astronomy and Astrophysics Encyclopedia,  S.
     Maran, ed., 	Van Nostrand Reinhold New York. p303
\ref Balasubrahmanyan, V. K., Boldt, E., \& Palmeira, R.,
\ref \ \ \ --- 1965, Phys. Rev. 140, B1157; 1967, J. Geophys. Res., 72, 27
\ref Balasubrahmanyan, V. K., Boldt, E., Palmeira, R. \& Sandri, G., 1968, Can. 	J. 	Phys., 46, S633
\ref Barcons, X. et al 1995, ApJ, 455, 480
\ref Bird, D. J. et al. 1995, ApJ, 441,144
\ref Blandford, R. D. \& Znajek, R., 1977,  MNRAS, 179, 433
\ref Bleach, R., Boldt, E., Holt, S., Schwartz, D. \& Serlemitsos, P., 1972, ApJ, 174, 	L101
\ref Bloemen, H., 1998, this workshop  
\ref Boldt, E., 1974, in High Energy Particles and Quanta in Astrophysics, eds F. McDonald \& C. Fichtel, MIT Press, Cambridge MA, Chapter  VIII
\ref Boldt, E., 1987, Physics Reports, 146, 215
\ref Boldt, E. \& Serlemitsos, P. 1969, ApJ, 157, 557
\ref Boldt, E. \& Leiter, D., 1995, Nuclear Physics B, 38, 440
\ref Boldt, E. \& Ghosh, P. 1998, MNRAS, submitted 
\ref Boyle, B. et al. 1998 MNRAS, 296, 1 (astro-ph/9710002)
\ref Chokshi, A. \& Turner, E. L., 1992 MNRAS, 259, 421
\ref Chung, D., Farrar, G. \& Kolb, E., 1998, Phys. Rev. D, 57, 4606
\ref Comastri, A. et al.  1995, A \& A , 296, 1
\ref Costa, E. et al. 1997, Nature, 387, 783
\ref Cronin, J., 1997, in Unsolved Problems in Astrophysics, eds Bahcall, J. N. \& Ostriker, J. P., 	Princeton University Press, p. 325
\ref Diehl, R. et al. 1995, A\&A, 298, 445
\ref Efstathiou, G., Ellis, R. \& Peterson, B. 1988 MNRAS, 323, 431
\ref Elbert, J. A. \& Sommers, P. J., 1995, ApJ, 441, 151
\ref Fabian, A. C. \& Rees, M. J., 1995, MNRAS, 277, L55
\ref Fabian, A. C. \& Barcons, X., 1992, ARA\&A, 30, 429
\ref Fabian, A. C. \& Canizares, C. R., 1988, Nat, 333, 829
\ref Farrar, G. \& Biermann, P. 1998, Phys. Rev., astro-ph/9806242
\ref Fichtel, C. E. et al. 1993, A\&A Supp. Ser., 97, 13
\ref Fisk, L., Kozlovsky, B. \& Ramaty, R., 1974, ApJ, 190, L35
\ref Friedlander, M., 1989, Cosmic Rays, Harvard U. Press, Cambridge
\ref Ghosh, P. \& Abramowicz, M. A., 1997, MNRAS, 292, 887
\ref Gloeckler, G. \& Jokippi, J. R., 1967, ApJ, 148, L41
\ref Greisen, K., 1966, Phys. Rev. Lett., 16, 748
\ref Hunter, S. et al. 1997, ApJ, 481, 205
\ref Jokippi, J. R. \& McDonald, F. B., 1995, Scientific American, 272, 58
\ref Kaaret, P., Ford, E. \& Chen, K., 1997, 480, L27
\ref Kinsey, J., 1970, Phys. Rev. Letters, 24, 246
\ref Kormendy, J. \& Richstone, D. O., 1995, ARA\&A, 33, 581
\ref Kormendy, J.,  et al., 1997, ApJ, 482, L139
\ref Koyama, K. et al. 1995. Nature, 378, 255
\ref Kozlovsky, B., Ramaty, R. \& Lingenfelter, R., 1997, ApJ, 484, 286
\ref Kuz'min, V. \& Tkachev, I., 1998  Pis'ma Eksp. Teor. Fiz., 68 (hep-ph/9802304)
\ref Lingenfelter, R.,  Ramaty, R. \& Kozlovsky, B.,1998, ApJ Letters, 500, L53 
\ref Loewenstein, M., Hayashida, K., Toneri, T., and Davis, D. S., 1998, ApJ, 497, 681
\ref Macdonald, D. \& Thorne, K., 1982, MNRAS, 198, 383
\ref Madau, P., Ghuisellini, G. and Fabian, A.C. 1994, MNRAS, 270, L17
\ref Magorrian, J. et al. 1998, AJ, 115, 2528  astro-ph/9708072
\ref Mahadevan, R., 1997, ApJ, 477, 585
\ref Mahoney, W., Ling, J., Wheaton, W. \& Jacobson, A.,  1984, ApJ, 286, 578
\ref McDonald, F. B. \& Ludwig, G. H., 1964, Phys. Rev. Letters, 13, 783
\ref McDonald, 1998 personal communication
\ref Metzger, M. R. et al. 1997, Nature, 387, 878
\ref Naranan, R., 1997, in Unsolved Problems in Astrophysics, eds Bahcall, J. N.  \& 	Ostriker, 	J.P., Princeton University Press, p. 301
\ref Naya, J. et al., 1996, Nature, 384, 44
\ref Padovani, P., Burg, R. \& Edelson, R., 1990, ApJ, 353, 438
\ref Parker, E. N., 1963, Interplanetary Dynamical Processes, John Wiley \& Sons, NY, 	Chapter XII
\ref Pierce, T. \& Blann, M. 1968, Phys. Rev., 173, 390
\ref Pohl, M., 1998, A\&A, in press
\ref Pravdo, S. \& Boldt, E., 1975, ApJ 200, 727
\ref Rees, M., 1997, in Unsolved Problems in Astrophysics, eds Bahcall, J. N. \& Ostriker, J. P., Princeton University Press, p. 181
\ref Reynolds, R. J., 1984, ApJ, 282, 191
\ref Richstone, D. et al. 1998, Nature, in press
\ref Rossi, B. 1964, Cosmic Rays, McGraw-Hill, New York
\ref Schmidt, M. 1978, Physica Scripta 17, 135
\ref Shklovskii, I., Moroz, V. \& Kurt, V., 1960, Astron. Zh., 37, 931
\ref Sigl, S., Lee, S.,  Schramm, D. \& Bhattacharjee, P., 1995, Sci, 270, 1977
\ref Simpson, J. A.  \& Connell, J. J., 1998, ApJ, 497, L85
\ref Small, T. A.\& Blandford, R. D., 1992, MNRAS, 259, 725
\ref Snowden, S. et al., 1997, ApJ, 485, 125
\ref Snowden, S., 1998, personal communication
\ref Soltan, A., 1982, MNRAS, 200, 115
\ref Sreekumar, P. et al. 1998, ApJ, 494, 523
\ref Streitmatter, R, 1998, in Proceedings of ``Workshop on Observing Giant Cosmic Ray Air Showers
from $>10^{20}$ eV Particles from  Space," eds J. Krizmanic \& J. Ormes, AIP Conf. Proc. 433, p.95
\ref Strong, A. W. et al. 1994, A \& A, 292, 82
\ref Tatischeff, V., Ramaty, R. \& Kozlovsky, B., 1998, ApJ, 504, 874 
\ref Thorne, K., 1974, ApJ, 191, 507
\ref Valinia, A. \& Marshall, F., 1998, ApJ, 505, 134 
\ref Van Allen, J. A., Ludwig, G. H., Ray, E. C. and McIlwain, C. E., 1958, Trans. 	AGU, 	39, 767
\ref von Ballmoos, P., 1995, Exp. Astron., 6, 85
\ref Zatsepen, G. T. \& Kuz'min, V. A., 1966, Pis'ma Eksp. Teor. Fiz., 4,114 [Sov Phys. JETP Lett. (Engl. Transl.) 4, 78]
\ref Zhang, W., Strohmeyer, T. \& Swank, J., 1997, ApJ, 482, L167
}                      

\end{document}